\documentclass[preprint,showpacs,preprintnumbers,amsmath,amssymb,natbib]{revtex4}

\usepackage{graphicx}
\usepackage{epsfig}		
\usepackage{dcolumn}
\usepackage{bm}
\def\0{\mbox{\tiny $0$}}
\def\1{\mbox{\tiny $1$}}
\def\2{\mbox{\tiny $2$}}
\def\3{\mbox{\tiny $3$}}
\def\4{\mbox{\tiny $4$}}
\def\5{\mbox{\tiny $5$}}
\def\6{\mbox{\tiny $6$}}
\def\7{\mbox{\tiny $7$}}
\def\8{\mbox{\tiny $8$}}
\def\9{\mbox{\tiny $9$}}

\def\f14{\mbox{\tiny $\frac{1}{4}$}}

\def\L{\mbox{\tiny $L$}}

\def\R{\mbox{\tiny $R$}}

\def\A{\mbox{\tiny $A$}}
\def\D{\mbox{\tiny $D$}}
\def\E{\mbox{\tiny $E$}}

\def\s{\mbox{\tiny $s$}}
\def\sss{\mbox{\tiny $S$}}

\def\j{\mbox{\tiny $j$}}
\def\mi{\mbox{\tiny $-$}}

\def\pl{\mbox{\tiny $+$}}
\def\al{\mbox{\tiny $\alpha$}}

\def\bb#1{\mbox{\footnotesize $(#1)$}}

\begin{document}

\title{Naturalness and stability of the generalized Chaplygin gas in the {\em seesaw cosmon} scenario}

\author{A. E. Bernardini}
\email{alexeb@ufscar.br, alexeb@ifi.unicamp.br}
\affiliation{Departamento de F\'{\i}sica, Universidade Federal de S\~ao Carlos, PO Box 676, 13565-905, S\~ao Carlos, SP, Brasil}
\altaffiliation{Also at Instituto Superior T\'ecnico, Departamento de F\'{\i}sica, Av. Rovisco Pais, 1, 1049-001, Lisboa, Portugal, as a visiting scholar}
\author{O. Bertolami}
\email{orfeu@cosmos.ist.utl.pt}
\affiliation{Instituto Superior T\'ecnico, Departamento de F\'{\i}sica, Avenida Rovisco Pais, 1, 1049-001, Lisboa, Portugal}
\altaffiliation[Also at]{~Instituto de Plasmas e Fus\~{a}o Nuclear, IST, Lisbon}

\date{\today}

\begin{abstract}
\renewcommand{\baselinestretch}{1.2}
\normalsize The {\em seesaw} mechanism is conceived on the basis that a mass scale, $\xi$, and a dimensionless scale, $s$, can be fine-tuned in order to control the dynamics of active and sterile neutrinos through {\em cosmon}-type equations of motion: the {\em seesaw cosmon} equations.
This allows for sterile neutrinos to be a dark matter candidate.
In this scenario, the dynamical masses and energy densities of active and sterile neutrinos can be consistently embedded into the generalized Chaplygin gas (GCG), the unified dark sector model.
In addition, dark matter adiabatically coupled to dark energy allows for a natural decoupling of the (active) mass varying neutrino (MaVaN) component from the dark sector.
Thus MaVaN's turn into a secondary effect.
Through the scale parameters, $\xi$ and $s$, the proposed scenario allows for a convergence among three distinct frameworks: the {\em cosmon} scenario, the {\em seesaw} mechanism for mass generation and the GCG model.
It is found that the equation of state of the perturbations is the very one of the GCG background cosmology so that all the results from this approach are maintained, being smoothly modified by active neutrinos.
Constrained by the {\em seesaw} relations, it is shown that the mass varying mechanism is responsible for the stability against linear perturbations and is indirectly related to the late time cosmological acceleration.
\end{abstract}

\date{\today}
\maketitle

\section{Introduction}

Theoretical efforts to interpret the observational data and to understand the nature of the dark sector do necessarily involve a vivid interplay between general relativity, astrophysics and particle physics.
Since the simplest solution to account for the late time accelerated expansion of the Universe, the one given in terms of a tiny positive cosmological constant, is plagued with conceptual problems, one has been compelled to examine different solutions \cite{Ame02,Kam02,Bil02,Ber02,Cal03,Mot04,Bro06A}.

Motivated by the high energy physics, an interesting alternative for obtaining the necessary negative pressure to account for the accelerated expansion involves the dynamics of a scalar field, $\phi$, evolving slowly down its potential $V\bb{\phi}$ \cite{Pee87,Rat87}.
These models assume that the vacuum energy can vary; a feature discussed much earlier \cite{Bro33}.
Other alternatives include $k$-essence \cite{Chi00,Arm01}, phantom energy models \cite{Sch01,Car03}, {\em cosmon} fields \cite{Wet88,Wet94}, and also several modifications of gravity \cite{Def02,Car04,Ama06}.

A challenging related issue concerns models of mass varying particles \cite{Gu03,Far04,Bja08}.
These are coupled to a light scalar field that drives their energy through their dynamical mass.
The idea of this mass varying mechanism \cite{Far04,Bro06A,Bja08} is to introduce a coupling between a relic particle, usually neutrinos, and the dark sector: dark energy or dark matter separately, or all together \cite{Ber08A,Ber08B}.
Such models admit an adiabatic regime in which the scalar field, usually related with dark energy, rolls down the minimum of its effective potential, which is set by the dark matter dynamical mass.
As a direct consequence of this new interaction, the particle mass is altered by the dynamics of the scalar field.

Due to phenomenological reasons, one still expects a small contribution from neutrinos to the cosmic dynamics.
In fact, it is well-known that the active neutrino masses are tiny as compared to the masses of the charged fermions.
The smallness of the neutrino masses is usually understood in terms of the {\em seesaw} mechanism in extensions of the standard model (SM) of the electroweak (EW) interactions.
The EW interactions involve only {\em left} handed neutrinos such that no renormalizable mass term for them is compatible with the SM gauge symmetry $SU(2)_{L}\otimes U(1)_{Y}$.
Once one assumes the conservation of baryon and lepton numbers, the {\em seesaw} mechanism admits neutrino masses from dimension five operators.
Neutrino masses should then involve two powers of the vacuum expectation value of the Higgs doublet.
These masses are suppressed by the inverse power of a large mass scale, $M$, of a {\em right} handed Majorana neutrino.
This super-heavy Majorana neutrino is associated to lepton number violating effects in extensions of the SM.

Assuming that the sterile neutrino mass, $\mathcal{M}\equiv \mathcal{M}\bb{\phi}$, exhibits a dynamical behavior driven by the scalar field, $\phi$, then the sterile neutrino becomes an interesting candidate for the aforementioned mass varying dark matter.
This, through the {\em seesaw} mechanism, gives origin to a mass varying neutrino (MaVaN) \cite{Gu03,Far04,Bja08} component corresponding to active neutrinos that exhibit secondary mass effects due to the {\em seesaw} mechanism.
In models of coupled dark energy, in which the scalar field couples to other matter components, it is natural to expect a coupling to active and to sterile neutrinos.

In this work, one explores the consequences of such a coupling.
This leads to cosmological scaling solutions where dark energy tracks the evolution of matter and/or radiation.
For the present cosmological epoch, it is predicted that the energy densities of dark energy and matter are of the same order of magnitude.
If the tracking can be altered by the growing mass of neutrinos, such that they become nonrelativistic (NR) at low redshift,
one can interestingly match the observed universe.

One should notice that mass varying dark matter is unusual in the formulation of the MaVaN models.
The unification of dark energy and dark matter naturally offers this possibility.
The generalized Chaplygin gas (GCG) is particularly relevant in this respect \cite{Kam02,Bil02,Ber02} as it is shown to be consistent with the observational constraints from CMB \cite{Ber03}, supernova \cite{Sup1,Ber04,Ber05}, gravitational lensing surveys \cite{Ber03B}, and gamma ray bursts \cite{Ber06B}.

Once one sets the mass dependence on $\phi$, the dark sector scalar field, one notices that the cosmological evolution of the unified fluid composed by mass varying dark matter and evolving dark energy has a dynamics similar to that of the {\em cosmon} field \cite{Wet94,Wet02,Wet08}.
Thus, at least partially, it turns out that the mass varying mechanism is the main agent of stability and of the cosmic acceleration.
Any cosmological fluid which effectively reproduces the effects of the GCG naturally offers analogous possibilities \cite{Ber09}.

Assuming that the scalar field drives the behavior of the masses of active and sterile neutrinos, the mass varying mechanism and the conditions for a stable cosmological scenario naturally emerges in the context of the GCC model.
Remarkably, the {\em seesaw} masses and energy densities of active and sterile neutrinos can be consistently embedded in the GCG scenario without any additional assumption.
Such a connection is mediated by a mass scale, $\xi$, and a dimensionless scale, $s$, similarly to {\em cosmon}-type dynamical equations \cite{Wet94,Wet02,Wet08}.
These are dubbed as {\em seesaw cosmon} equations.
This procedure gives origin to a remarkable convergence of three distinct frameworks: the {\em cosmon}-like dynamics, the {\em seesaw} mechanism for mass generation and the GCG scenario.
The equation of state of the perturbations is the very one of the background cosmology so that all effective results arising from the GCG model are maintained, although modified by neutrinos in a quite subtle way.

This work is organized as follows.
The decoupling mechanism for active and sterile neutrinos is described in section II.
This is obtained through the coupling of a scalar field in the {\em seesaw} relations, which leads to the dynamical properties ensued by the mass varying mechanism.
In section III, the interplay with the GCG model is discussed.
In section IV, the main properties of a unified treatment of dark matter and dark energy in a dark energy scenario for a $\Lambda$-like equation of state, $p = - \rho$, is reviewed.
This leads to a dynamical mass prescription different from the one obtained through the simplified version of the {\em seesaw} mechanism explored in section III.
Specifical mass dependencies on the scalar field for which the mass varying dark matter results in an {\em
effective} GCG model are discussed.
Finally, the results for energy densities, growing neutrino mass and stability conditions from a positive squared speed of sound $c_{s}^{\2}$ are discussed in section V.
Three cosmological scenarios are examined, actually associated to three different growing mass relations: $\mu\bb{\phi} = \phi \exp{\left[-3\bb{\alpha + 1}\phi/2\right]}$ (Case 01),  $\mu\bb{\phi} = \exp{\left[-3\bb{\alpha + 1}\phi/2\right]}$ (Case 02) and $\mu\bb{\phi} = \tanh{\left[3\bb{\alpha + 1}\phi/2\right]}^{\frac{\2\alpha}{\alpha + 1}}\exp{\left[-3\bb{\alpha + 1}\phi\right]}$ (Case 03).
The obtained results indicate that the proposed approach is quite appealing as it unifies neutrinos, dark matter and dark energy.
In section VI, the main implications of the developed scenario are summarized.

\section{The {\em seesaw} mechanism for MaVaN's}

The connection of mass varying dark matter with neutrinos does provided, as will be seen, interesting constraints on the neutrino masses, on the dark energy density, and on the equations of state and the stability conditions.
This can be understand through the equations arising from the Lagrangian densities of active ($A$) and sterile ($S$) neutrinos, $\psi_{\A,\sss}$,
\begin{equation}
\mathcal{L}_{\A,\sss} = i\bar{\psi}_{\A,\sss}\gamma_{\mu}\partial^{\mu}\psi_{\A,\sss} + k_{\A,\sss}\bar{\psi}_{\A,\sss}\psi_{\A,\sss},
\label{aprl00A}
\end{equation}
where two mass scales, $k_{\A} = \mu$ and $k_{\sss} = \mathcal{M}$, have been introduced.
The {\em seesaw} mechanism suggests that $\mu$ is small, while $\mathcal{M}$ should be large:
\begin{equation}
-\mu = (M/2)[1 - \sqrt{1 + 4 (m/M)^{\2}}] ~~\mbox{and}~~\mathcal{M} = (M/2)[1 + \sqrt{1 + 4 (m/M)^{\2}}].
\label{aprl02}
\end{equation}
These states correspond to the eigenvalues of the mass matrix $\left[\begin{array}{cc} 0 & m\\ m & M\end{array}\right]$ written in the orthogonal basis of chiral {\em left}- and {\em right}-handed neutrinos, $\nu_{\L,\R}$, related with the matter fields, $\psi_{\A,\sss}$, by
\begin{equation}
\psi_{\A} = \sqrt{s^{\2} + 1} (s \nu_{\L} - \nu_{\R})~~\mbox{and}~~
\psi_{\sss} = \sqrt{s^{\2} + 1} ( \nu_{\L} + s \nu_{\R}),
\label{aprl00B}
\end{equation}
where the dimensionless quantity $s = \sqrt{\mu/\mathcal{M}}$ has been introduced.
It follows that $\mathcal{L}_{\A} + \mathcal{L}_{\sss} = \mathcal{L}_{\L} + \mathcal{L}_{\R} + \mathcal{L}_{\L\R}$, where $\mathcal{L}_{\L,\R}$ correspond to the kinetic terms and $\mathcal{L}_{\L\R}$ yields the mass mixing terms.
The equivalence between the stress-energy tensor in the chiral basis and the matter field basis follows from the relationships for the energy density $\rho_{\A} + \rho_{\sss} = \rho_{\L} + \rho_{\R}$ and for the pressure $p_{\A} + p_{\sss} = p_{\L} + p_{\R}$.
After introducing an auxiliary mass scale $\xi = \sqrt{\mu \mathcal{M}} \equiv m$, one can define two energy scales, $\rho_{\1} = (\rho_{\sss} + \rho_{\A})/2$ and $\rho_{\2} = (\rho_{\sss} - \rho_{\A})/2$ which evolve as reciprocally coupled {\em cosmon}-type equations.
In a FRW universe it corresponds to
\begin{equation}
\dot{\rho}_{\1} + 3 H (\rho_{\1} + p_{\1}) - \dot{\phi}\frac{\mbox{d} \xi}{\mbox{d} \phi} \frac{\partial \rho_{\1}}{\partial \xi} + \dot{\phi}\frac{\mbox{d} s}{\mbox{d} \phi} \frac{\partial \rho_{\2}}{\partial s} = 0,
\end{equation}
and
\begin{equation}
\dot{\rho}_{\2} + 3 H (\rho_{\2} + p_{\2}) - \dot{\phi}\frac{\mbox{d} \xi}{\mbox{d} \phi} \frac{\partial \rho_{\2}}{\partial \xi} + \dot{\phi}\frac{\mbox{d} s}{\mbox{d} \phi} \frac{\partial \rho_{\1}}{\partial s} = 0,
\label{aprl02A}
\end{equation}
where $H = \dot{a}/{a}$ is the expansion rate of the universe and the {\em overdot} denotes differentiation with respect to time ($^{\cdot}\, \equiv\, d/dt$).
The third terms in the above Eqs. correspond to a mass varying term (see Eq.~(\ref{gcg02}) in the Appendix), while the last ones are associated to the exchange of energy due to the non-adiabatical behavior of energy densities $1$ and $2$.
Notice that there is no sense in defining a Lagrangian density for the component $\rho_{\2}$, which does not correspond to an observable energy density scale.
The $\rho_{\1}$ component can be identified to matter fields, while $\rho_{\2}$ is an auxiliary energy density which measures the coupling between the physical observables.
In fact, both energy densities, $1$ and $2$, are driven by {\em cosmon}-type equations.
The {\em cosmon} signature is revealed by the dependence of the scales $\xi$ and $s$ on the value of a slowly varying classical scalar field $\phi$, the {\em seesaw cosmon} field \cite{Wet94}.
Since the scalar field depends on the scale factor $a$, $\phi \equiv \phi\bb{a}$, the {\em seesaw} mass terms get transformed into dynamical quantities, $\mu\bb{\phi}$ and $\mathcal{M}\bb{\phi}$.
After combining Eqs.~(\ref{aprl02}), and observing that,
\begin{equation}
\rho_{\A}\bb{a, \xi, s} = \rho_{\A}\bb{a, \xi s},
\label{aprl023A}
\end{equation}
and
\begin{equation}
\rho_{\sss}\bb{a, \xi, s} = \rho_{\sss}\bb{a, \xi/s},
\label{aprl023}
\end{equation}
it is easy to identify the evolution of these with the evolution of active and sterile energy densities, $\rho_{\A}$ and $\rho_{\sss}$, by means of decoupled equations
\begin{equation}
\dot{\rho_{\A}} + 3 H (\rho_{\A} + p_{\A}) - \dot{\phi}\frac{\mbox{d} \mu}{\mbox{d} \phi} \frac{\partial \rho_{\A}}{\partial \mu} = 0,
\label{aprl03A}
\end{equation}
and
\begin{equation}
\dot{\rho_{\sss}} + 3 H (\rho_{\sss} + p_{\sss}) - \dot{\phi}\frac{\mbox{d} \mathcal{M}}{\mbox{d} \phi} \frac{\partial \rho_{\sss}}{\partial \mathcal{M}} = 0,
\label{aprl03B}
\end{equation}
as
\begin{equation}
\frac{\mbox{d} \ln{\xi}}{\mbox{d} \phi} =
(1/2)\left(\frac{\mbox{d} \ln{\mu}}{\mbox{d} \phi} + \frac{\mbox{d} \ln{\mathcal{M}}}{\mbox{d} \phi}\right)\nonumber\\,
\end{equation}
and
\begin{equation}
\frac{\mbox{d} \ln{s}}{\mbox{d} \phi} =
(1/2)\left(\frac{\mbox{d} \ln{\mu}}{\mbox{d} \phi} - \frac{\mbox{d} \ln{\mathcal{M}}}{\mbox{d} \phi}\right)\nonumber\\.
\end{equation}
Thus, mass varying mechanism translates the dependence of the mass terms on the scale factor, $a$, i.e. $\mu\bb{a}$ and $\mathcal{M}\bb{a}$ (see the Appendix).
The coupling between relic particles and the scalar field as described by Eqs.~(\ref{aprl03A})-(\ref{aprl03B}) is relevant only for NR fluids .
Since the strength of the coupling is suppressed by the relativistic pressure, as long as particles are ultra-relativistic (UR), the active and sterile neutrino energy densities, $\rho_{\A}$ and $\rho_{\sss}$, decouple from each other and evolve adiabatically \cite{Bea08}, remaining coupled only to the scalar field.
That is, in the UR regimes, $\frac{\partial\rho}{\partial m}\sim\frac{\partial\rho}{\partial s}\propto (\rho- 3 p)\approx 0$.

In a previous work, it has been suggested that one could treat MaVaN's as a perturbative component derived from an unperturbed adiabatic energy density solution $\rho_{\phi}$ \cite{Ber08A,Ber08B}.
The above results provides a quantitative justification for that.
As can be seen, all the information from the dark sector (dark energy plus dark matter) acting on the (active) neutrino sector is carried out by the explicit dependence of $\mu \equiv\mu\bb{\phi}$.
From the cosmological point of view, it results in mass eigenstantes for active award sterile neutrinos that evolve separately, which is not the case for the coupled chiral eigenstates $\nu_{\L}$ and $\nu_{\R}$.
At primordial times, when $s^{\2} \sim 1$, such mass eigenstates are indistinguishable, and the chiral eigenstates are well-defined.
At late times they turn into sterile and active mass eigenstates, maintaining the identity of the flavour sectors.

In order to proceed, one identifies the large mass energy density component of the above equations, $\rho_{\sss}$, to the energy density of dark matter, then the {\em seesaw cosmon} framework provides the connection between dark matter and dark energy through the {\em cosmon} field equation \cite{Wet88,Wet94},
\begin{equation}
\dot{\rho}_{\phi} + 3 H (\rho_{\phi} + p_{\phi}) + \dot{\phi}\frac{\mbox{d} \mathcal{M}}{\mbox{d} \phi} \frac{\partial \rho_{\sss}}{\partial \mathcal{M}} = 0,
\label{aprl03C}
\end{equation}
originally written as
\begin{equation}
\ddot{\phi} + 3 H \dot{\phi} + \frac{\mbox{d} V\bb{\phi}}{\mbox{d} \phi} =  - \frac{\mbox{d} \mathcal{M}}{\mbox{d} \phi} \frac{\partial \rho_{\sss}}{\partial \mathcal{M}},
\label{gcg04}
\end{equation}
with the usual assignments for the $\rho_{\phi}$ and $p_{\phi}$ (c. f. Eqs.(\ref{pap01})).
Eqs.~(\ref{aprl03B}) - (\ref{aprl03C}) result in the adiabatic equation for the dark sector,
\begin{equation}
\dot{\rho}_{\D\sss} + 3 H (\rho_{\D\sss} + p_{\D\sss}) = 0,
\label{aprl04}
\end{equation}
with $\rho_{\D\sss} = \rho_{\phi} + \rho_{\sss} = \rho_{\phi} + \rho_{\1} + \rho_{\2}$ and $H^{\2} = \rho_{\D\sss}$ (with $H$ in units of $H_{\0}$ and $\rho_{\D\sss}$ in units of $\rho_{\mbox{\tiny Crit}} = 3 H^{\2}_{\0}/ 8 \pi G)$.
Despite the intrinsic dependence on $\phi$, the equation of motion for the dark sector is not modified by $\rho_{\A}$, the active neutrino energy density component.
The cosmological dependence of $\rho_{\A}$ on $\phi$ can be computed through Eq.~(\ref{aprl03A}) considering active neutrinos as a test fluid.

The phenomenological consistency of the proposed scenario can be assessed quantitatively expressing $\rho_{\A}$ and $\rho_{\sss}$ as energy densities of a degenerate fermion gas (DFG) at different relativistic regimes,
\small
\begin{eqnarray}
\rho_{\A}\bb{a} &=& (8 \pi^{\2})^{\mi\1}
\mu\bb{a}^{\4}\left[\eta\bb{a} (2 \eta\bb{a}^{\2} + 1)\sqrt{\eta\bb{a}^{\2} + 1} -
\mbox{arc}\sinh{(\eta\bb{a})}\right]\nonumber\\
\rho_{\sss}\bb{a} &=& (8 \pi^{\2})^{\mi\1}
\mu\bb{a}^{\4}\, s^{\mi\8}\left[\gamma s^{\2} \eta\bb{a} (2 \gamma^{\2} s^{\4} \eta\bb{a}^{\2} + 1)\sqrt{\gamma^{\2} s^{\4} \eta\bb{a}^{\2} + 1} -
\mbox{arc}\sinh{(\gamma s^{\2} \eta\bb{a})}\right],
\label{aprl04B}
\end{eqnarray}
\normalsize
where $\eta\bb{a} = (T_{\0}\,q_{\A})/(a\,\mu\bb{a})$ and the relation between the fluid thermodynamic regimes is parameterized by the coefficient $\gamma = q_{\sss}/q_{\A}$.
As discussed in the Appendix, the DFG prescription is suitable for describing the transition between UR and NR regimes.
In this case, the effects due to the coupled dark matter ($\mathcal{M}\bb{\phi}$) and dark energy ($\phi$) can be monitored through Eqs.~(\ref{aprl03A})-(\ref{aprl03C}).

\begin{figure}
\vspace{-2.5 cm}
\centerline{\psfig{file= 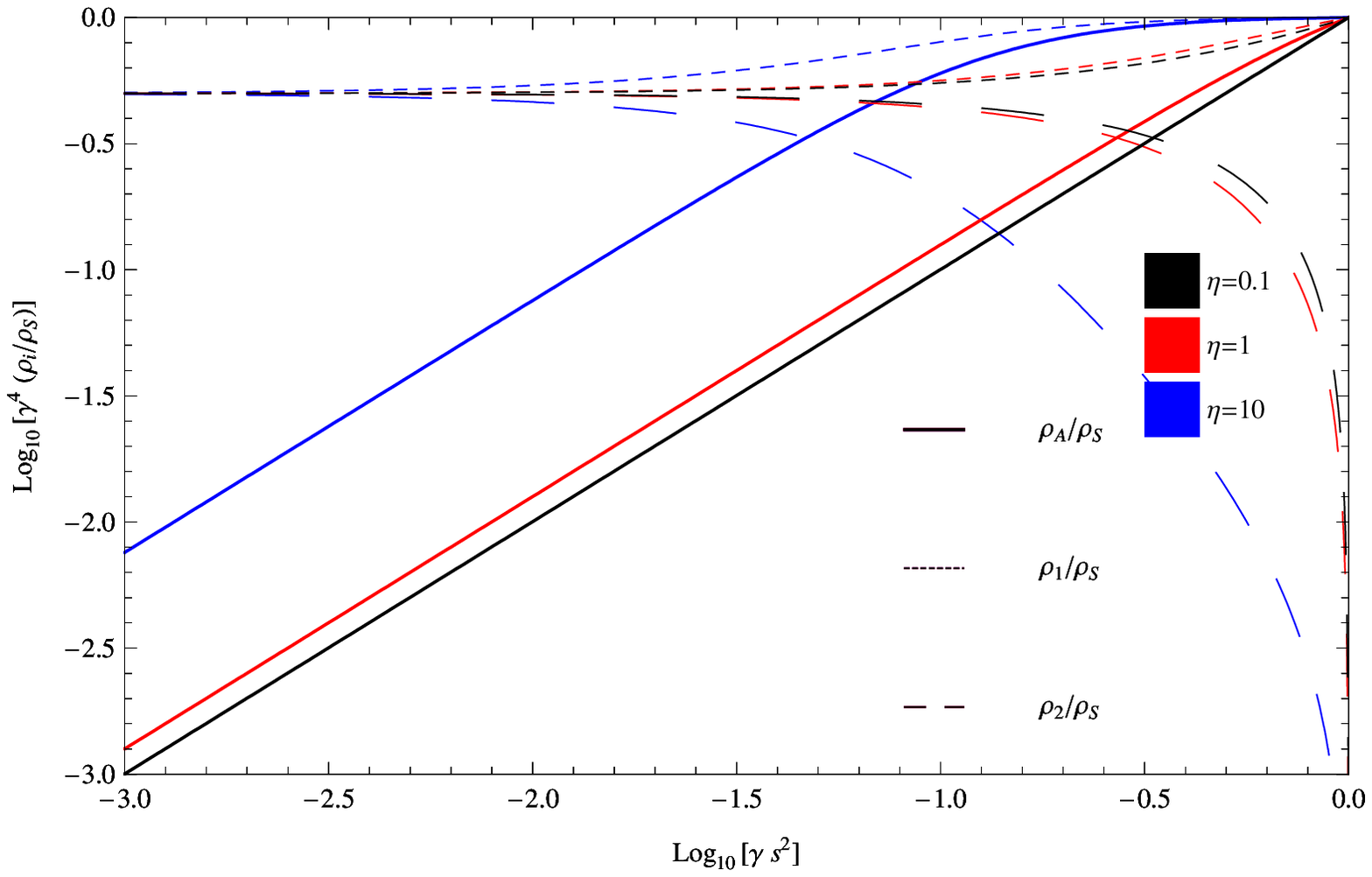,width=11 cm}}
\centerline{\psfig{file= 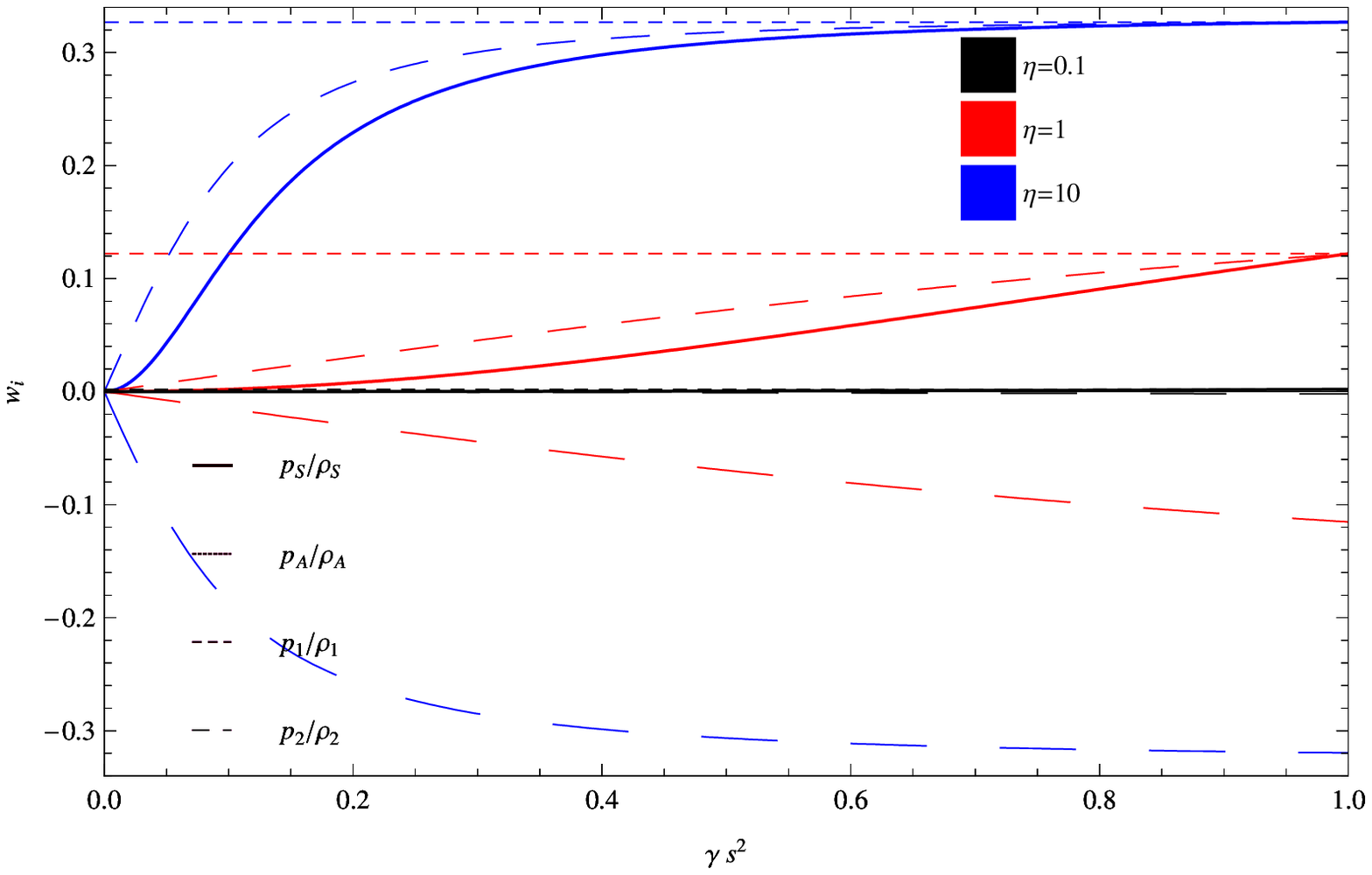,width=11 cm}}
\caption{\small
The correspondence between the energy density components and the ``modified'' {\em seesaw} scale parameter $\gamma s^{\2}$ (first plot), and the equation of state $w_{i} = p_{i}/\rho_{i}$.
Sterile and active neutrinos are assumed to behave like a DFG in NR ($\eta = 0.1$),  relativistic ($\eta = 1$) and  UR ($\eta = 10$) regimes.
For the NR limit of a DFG, on has $\rho_{\A}/\rho_{\sss} \sim \gamma^{\mi\3} s^{\2}$ (straight solid line in the plot).}
\label{PRL01}
\end{figure}

In the Fig.~\ref{PRL01} one can see the exact correspondence between the abovementioned energy densities and the ``modified'' scale parameter $\gamma s^{\2}$.
In the NR limit of a DFG, one has $\rho_{\A}/\rho_{\sss} \sim \gamma^{\mi\3} s^{\2}$.

The characteristic magnitude of the active neutrino masses involves an appropriate combination of dimensionless Yukawa couplings, $Y_{\j}$, $\mu_{\j} \sim m_{\j}^{\2}/M$ with $m_{\j} \sim Y_{\j}\, v$ ($v\sim 2 \times 10^{\1\1}\,eV$).
Consistency with the observed flavour oscillations requires for the neutrino mass at least one neutrino to have $\mu_{\j}\gtrsim 0.05 \,eV$.
That is, $Y_{\j}$ of the order one implies an upper bound for the large mass scale $\mathcal{M} \lesssim 10^{\2\3}\,eV$, from which follows that $s^{\2} \gtrsim 10^{\mi\2\4}$.
Given that the present value of the ratio $\rho_{\A}/\rho_{\sss}$ is, for phenomenological reasons, of $\mathcal{O}(10^{\mi\2})$, a DFG of active neutrinos, at least approximately in the NR regime, leads to $\gamma^{\mi\3}\sim 10^{\2\2}$, and hence, two widely different momentum scales for active and sterile neutrinos, $q_{\sss}/q_{\A} \sim 10^{\mi\7}$.
Thus, sterile neutrinos behave like ultra cold dark matter (CDM).
In addition, if one assumes that $\rho_{\phi}$ could be parameterized by a $\lambda \phi^{\4}$ theory, or some type of quintessence potential, the adiabatic evolution of the scalar field allows for assigning a mass for the scalar field.
Such predictions for $m_{\phi}$ are consistent with the lower bound on the mass derived from assumptions about the adiabatic evolution of the scalar field since nucleosynthesis.
The scalar field mass should then be greater than the rate of expansion at nucleosynthesis, which is of order of $\sim 10^{\mi\1\6}\, eV$.
However, the conditions for the adiabatic regime for the light scalar field to be settled down at the minimum of its potential prior to nucleosynthesis are quite model dependent.

The above discussion shows that a considerable fine-tuning is required to generate the different scalar field mass, a issue that demands for an embedding of the model in a more encompassing framework, such as, for instance, the Minimal Supersymmetric SM with the addition of one singlet chiral superfield \cite{Bal07}.
Another possible avenue involves a unified picture of dark energy and dark matter where the former corresponds to an additional singlet scalar field which one identifies to the quintessence field \cite{Ber09C,Ber09D}, while the latter with the quantum excitations of this singlet, which can be coupled to the Higgs boson.

\section{The interplay with the GCG model}

One considers now some generic features of the proposed model and explores its connection with the GCG model.

The GCG model is characterized by an exotic equation of state \cite{Ber02,Ber03} given by
\begin{equation}
p = - A_{\s} \rho_{\0} \left(\frac{\rho_{\0}}{\rho}\right)^{\al},
\label{aprl05}
\end{equation}
where $A_{\s}$ and $\alpha$ are constants.
This equation of state can be obtained from a generalized Born-Infeld action \cite{Ber02}.
Several studies yield convincing evidence that the GCG scenario is a phenomenologically consistent approach to explain the accelerated expansion of the universe.
Inserting the above equation of state into the unperturbed energy conservation Eq.~(\ref{aprl04}), one obtains through a straightforward integration \cite{Ber02}
\begin{equation}
\rho = \rho_{\0} \left[A_{\s} + \frac{(1-A_{\s})}{\bb{a/a_{\0}}^{\3(\1+\alpha)}}\right]^{\1/(\1 \pl \al)},
\label{gcg21}
\end{equation}
and
\begin{equation}
p = - A_{\s} \rho_{\0} \left[A_{\s} + \frac{(1-A_{\s})}{\bb{a/a_{\0}}^{\3(\1+\alpha)}}\right]^{-\al/(\1 \pl \al)}.
\label{gcg22}
\end{equation}
One of the most striking features of the above equations is that the energy density interpolates between a dust dominated phase in the past, where $\rho \propto a^{-\3}$, and a de-Sitter phase, $\rho = -p$, at late times.
This evolution is ruled by the model parameters, $\alpha$ and $A_{\s}$, which are positive and smaller than unity.
Of course, $\alpha = 0$ corresponds to the $\Lambda$CDM model.
It sets $0 < \alpha \leq 1$.
For $A_{\s} =0$, GCG behaves like matter, whereas for $A_{\s} =1$, it behaves as a cosmological constant.
Hence, in order to consider it as unified model for dark matter and dark energy one has to exclude these two possibilities so that $A_{\s}$ must lie in the range $0 < A_{\s} < 1$.
This property makes the GCG model an interesting candidate for the unification of dark matter and dark energy, i. e. for the dark sector energy density $\rho_{\D\sss}$ of our proposal.

The GCG can be described by an underlying scalar field, $\phi$, which can be either real \cite{Kam02,Ber04} or complex \cite{Bil02,Ber02}.
In the former case, one can identify it with the {\em cosmon} field $\phi$ so that $\rho$ and $p$ are given in terms of $\phi$ by:
\begin{eqnarray}
\rho &=& \frac{1}{2}\dot{\phi}^{\2} + V\bb{\phi},\nonumber\\
p    &=& \frac{1}{2}\dot{\phi}^{\2} - V\bb{\phi}.
\label{pap01}
\end{eqnarray}
This allows for obtaining the dependence of the scalar field, $\phi$, on the scale factor, $a$, and the explicit dependence of $\rho$, $p$ and $V$ on $\phi$.
Following Ref.~\cite{Ber04}, one obtains through Eqs.~(\ref{gcg21})-(\ref{pap01})
\begin{equation}
\dot{\phi}^{\2}\bb{a} = \frac{\rho_{\0}(1 - A_{\s})}{\bb{a/a_{\0}}^{\3(\al\pl\1)}}
\left[A_{\s} + \frac{(1-A_{\s})}{\bb{a/a_{\0}}^{\3(\al\pl\1)}}\right]^{-\al/(\al \pl \1)},
\label{pap02}
\end{equation}
and assuming a flat universe described by Friedmann equation $H^{\2} = \rho$ (again with $H$ in units of $H_{\0}$ and $\rho$ in units of $\rho_{\mbox{\tiny Crit}} = 3 H^{\2}_{\0}/ 8 \pi G)$, one gets
\begin{equation}
\phi\bb{a} = - \frac{1}{3(\alpha + 1)}\ln{\left[\frac{\sqrt{1 - A_{\s}(1 - \bb{a/a_{\0}}^{\3(\al \pl \1)})} - \sqrt{1 - A_{\s}}}{\sqrt{1 - A_{\s}(1 - \bb{a/a_{\0}}^{\3(\al \pl \1)})} + \sqrt{1 - A_{\s}}}\right]},
\label{pap03}
\end{equation}
where it is assumed that
\begin{equation}
\phi_{\0} = \phi\bb{a = a_{\0}} = - \frac{1}{3(\alpha + 1)}\ln{\left[\frac{1 - \sqrt{1 - A_{\s}}}{1 + \sqrt{1 - A_{\s}}}\right]}.
\label{pap04}
\end{equation}
One then readily finds the scalar field potential,
\begin{equation}
V\bb{\phi} = \frac{1}{2}A_{\s}^{\frac{\1}{\1 \pl \al}}\rho_{\0}\left\{
\left[\cosh{\left(3\bb{\alpha + 1} \phi/2\right)}\right]^{\frac{\2}{\al \pl \1}}
+
\left[\cosh{\left(3\bb{\alpha + 1} \phi/2\right)}\right]^{-\frac{\2\al}{\al \pl \1}}
\right\}.
\label{pap05}
\end{equation}
Thus, the dynamics of the GCG scalar field is given by the evolution of $\phi$ on the above potential.

Turning now to the simplest version of the {\em seesaw} mechanism for which a single flavour neutrino mass is linearly related to the scalar field, $\phi$, i. e. $m\bb{\phi} \sim \phi$, after observing that the logarithm of the squared scale parameter $s^{\2}$ has an analytical structure similar to that of $\phi\bb{a}$,
\begin{equation}
\ln{(s^{\2})} =\ln{(\mu/\mathcal{M})} = \ln{\left[\frac{\sqrt{1 + 4 (m/M)^{\2}} - 1}{\sqrt{1 + 4 (m/M)^{\2}} + 1}\right]},
\label{aprl09}
\end{equation}
one can rewrite the auxiliary scale $m/M$ in terms of the GCG parameters $A_{\s}$ and $\alpha$, and on terms of the scale factor, $a$, as
\begin{equation}
\frac{m\bb{a}}{M\bb{a}} = \frac{m_{\0}}{M_{\0}}  a^{\frac{3\bb{\al + \1}}{2}} = \frac{1}{2} \sqrt{\frac{A_{\s}}{1-A_{\s}}} (a/a_{\0})^{\frac{3\bb{\al + \1}}{2}}
\label{aprl10}
\end{equation}
where $a_{\0}$ is determined by a phenomenological adjustment (it can be set equal to unity if one redefines $A_{\s}$ and $\rho_{\0}$ in Eq.~(\ref{gcg21})).
After a simple mathematical manipulation, one obtains
\begin{equation}
\ln{(s^{\2})} = \ln{(\mu/\mathcal{M})} = - 3\bb{\alpha + 1} \phi\bb{a},
\label{aprl09B}
\end{equation}
Eqs.~(\ref{aprl09})-(\ref{aprl09B}) yield the following relationships for active and sterile neutrino masses
\begin{equation}
\mu\bb{\phi} = \phi \exp{\left[-3\bb{\alpha + 1} \phi/2\right]} ~~~~ \mbox{and} ~~~~ \mathcal{M}\bb{\phi} = \phi \exp{\left[+3\bb{\alpha + 1} \phi/2\right]}.
\label{aprl11}
\end{equation}
The behavior set by Eqs.~(\ref{aprl10}) and (\ref{aprl11}) implies to the scale parameter $s$ an exponential dependence on $\phi$, $s = \exp{\left[-3\bb{\alpha + 1} \phi/2\right]}$, corresponding to an exponential divergency between mass scales.
This is consistent with several classes of quintessence models \cite{Bar99}.
Naturally, once the prescription for masses and couplings is known, the behavior of the neutrinos can be understood through Eqs.~(\ref{aprl03A}) and (\ref{aprl03B}).

It is evident that the procedure above described is fairly general.
The form of $M\bb{\phi}$ and modifications on the equation of state for dark energy can lead to quite different scenarios.
For instance, it also admits constant mass dark matter, which should result in neutrino masses given by $\mu\bb{a} \propto \exp{\left[-3\bb{\alpha + 1}\,\phi\bb{a}\right]}$, or even a constant mass scale $\xi = m$, as will be discussed in the analysis of the following section.

\section{Mass varying dark matter from an effective GCG}

Aiming to obtain a deeper understanding of the proposal of the previous section, one shows that the coupling of dark matter to dark energy through the mass varying mechanism in the {\em seesaw cosmon} scenario can be indeed matched with the GCG model.
In order to perform that, one assumes that the GCG equation of state Eq.~(\ref{aprl05}) just describes the behavior of a fluid with energy density $\rho_{\D\sss}$.

From Eqs.(\ref{aprl03B}) and (\ref{aprl03C}), one sees that if the {\em seesaw cosmon} field obeys the following equation of state
\begin{equation}
\rho_{\phi}\bb{\phi} = - p_{\phi}\bb{\phi} = U\bb{\phi} = \left[A_{s}\cosh{\left(\frac{3\bb{\alpha + 1}\phi}{2}\right)}\right]^{\frac{\2 \al}{1 \pl \al}},
\label{pap08}
\end{equation}
thus the problem consists in obtaining the relationship between the scalar potential $U\bb{\phi}$ and the variable mass $\mathcal{M}\bb{\phi}$ which satisfies the equation
\begin{equation}
\frac{\mbox{d} U\bb{\phi}}{\mbox{d}{\phi}} +
\frac{\partial \rho_{\sss}}{\partial \mathcal{M}} \frac{\mbox{d} \mathcal{M}\bb{\phi}}{\mbox{d}\phi} = \frac{\mbox{d} U_{\mbox{\tiny Eff}}\bb{\phi}}{\mbox{d}{\phi}} = 0,
\label{pap09}
\end{equation}
a stationary condition directly obtained from Eq.~(\ref{aprl03C}).
The effective potential $U_{\mbox{\tiny Eff}}\bb{\phi}$ that sets the evolution of the scalar field has two terms: the first one arising from the original quintessence potential $U\bb{\phi}$, and the second one due to a coupling to the dynamical mass $\mathcal{M}\bb{\phi}$.
For suitable choices of $U\bb{\phi}$ and for couplings satisfying Eq.~(\ref{pap09}), the competition between these terms leads to a minimum of the effective potential.
In the {\em quasi}-static regime, it is possible for the field to adjust itself to the minimum of the potential in an adiabatic way.
The timescale for that must be short when compared to the timescale over which the background density is changing.
In this regime, matter and scalar field are tightly coupled and evolve effectively as a single fluid.
In the proposed approach, once assuming the equation of state $p_{\phi} = -\rho_{\phi}$, the described behavior is rather natural.
Eq.~(\ref{pap08}) leads to $\rho_{\D\sss} + p_{\D\sss} = \rho_{\sss} + p_{\sss}$ which, in the CDM limit, gives $p_{\sss} = 0$ and $\rho_{\D\sss} + p_{\D\sss} = \rho_{\sss}\bb{a} = \mathcal{M}\bb{a} \, n\bb{a}$, where $n\bb{a}$ is the particle number density.
Since the dependence of $\mathcal{M}$ on $a$ is mediated by $\phi\bb{a}$, i. e. $\mathcal{M}\bb{a} = \mathcal{M}\bb{\phi\bb{a}}$,
from Eqs.~(\ref{gcg21}), (\ref{gcg22}) and (\ref{pap03}), it follows after some mathematical manipulations that
\begin{equation}
\mathcal{M}\bb{\phi} = \mathcal{M}_{\0}
\left[\frac{\tanh{\left(3\bb{\alpha + 1}\frac{\phi}{2}\right)}}{\tanh{\left(3\bb{\alpha + 1}\frac{\phi_{\0}}{2}\right)}}\right]^{\frac{ \2 \al}{1 \pl \al}},
\label{pap11}
\end{equation}
which is consistent with Eq.~(\ref{pap09}).
Hence, one sees that the adequacy to the adiabatic regime is conditioned by the mass varying behavior.

In a previous study \cite{Ber09B}, it has shown how the behavior of $\mathcal{M}\bb{\phi}$ and $U\bb{\phi}$ differs from the one of  $\phi\bb{a}$ and $U\bb{\phi}$ in the GCG model, and how the composed fluid deviates from the GCG scenario.
It has been assumed that the mass varying dark matter behaves like a DFG in the relativistic regime (hot dark matter (HDM)) and in the NR regime (CDM).
For the mass varying CDM coupled to dark energy with $p_{\phi} = -\rho_{\phi}$, the GCG leads to similar predictions for $w = p_{\phi} / \rho_{\phi}$, independently of the scale parameter $a$.
The same is not true for HDM as the GCG-like behavior holds just close to the present ($a \sim a_{\0}$).

\section{Energy density, neutrino mass and stability}

As mentioned, the evolution of matter components are constrained by several factors.
The precise peak location of the CMB anisotropies, the change in the growth of cosmic structures, and the properties of nonlinear structures which provide detailed information to test such models.
In here a preliminary analysis is performed to compare neutrino masses and the corresponding conditions for stability for three cases of neutrino dynamical masses.
Due to previous arguments, these cases take place in the context of the GCG model.
The first two cases correspond to a neutrino mass dependence on the {\em cosmon} field.
For the first case, the off-diagonal mass matrix elements, the Dirac mass terms, $m\bb{\phi}$, have a linear dependence on $\phi$, $m\bb{\phi}\sim \phi$, which results in a tiny mass eigenvalue $\mu\bb{\phi} = \phi \exp{\left[-3\bb{\alpha + 1}\phi/2\right]}$.
For the second case, the off-diagonal mass matrix elements are constants, $m\bb{\phi}\sim const.$, and the dynamical masses are generated by the dependence on the Majorana mass term $M\bb{\phi}$, which results in $\mu\bb{\phi} = \exp{\left[-3\bb{\alpha + 1}\phi/2\right]}$.
The third case describes a sterile neutrino with a dynamical mass evolving like the GCG with mass varying dark matter coupled to dark energy with a equation of state, $p_{\D\E} = - \rho_{\D\E}$.
Since the prescription for decoupling mass varying dark matter from the effective GCG leads to sterile neutrino mass given by Eq.~(\ref{pap11}), the active neutrino mass should be given by $\mu\bb{\phi} = \tanh{\left[3\bb{\alpha + 1}\phi/2\right]}^{\frac{\2\alpha}{\alpha + 1}} \exp{\left[-3\bb{\alpha + 1}\phi\right]}$.
Keeping in mind the connection established by the {\em seesaw} mechanism for the third case, thus the mass matrix term should be given by
$m\bb{\phi} = \tanh{\left[3\bb{\alpha + 1}\phi/2\right]}^{\frac{\2\alpha}{\alpha + 1}} \exp{\left[-3\bb{\alpha + 1}\phi/2\right]}$.

\begin{figure}
\centerline{\psfig{file= 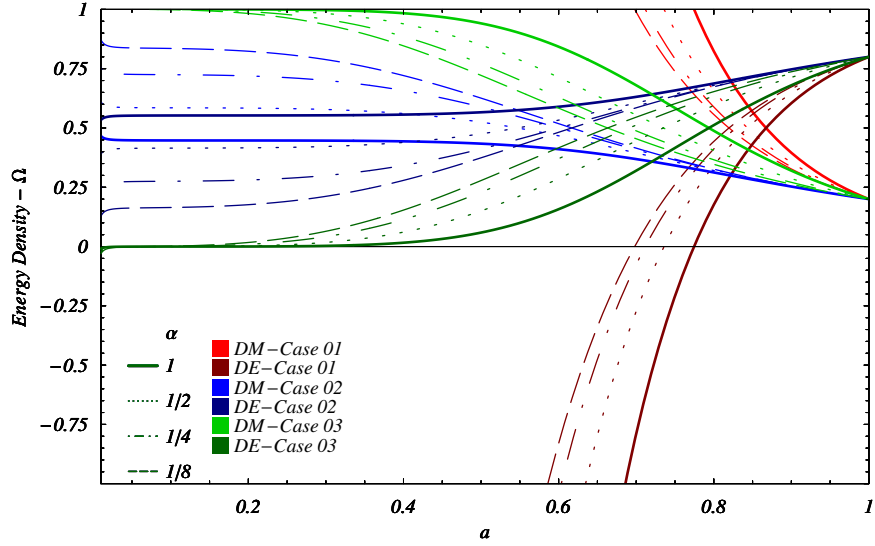,width=13 cm}}
\caption{\small Energy densities $\Omega = \rho\bb{a}/\rho_{\0}$ in terms of the scale factor, $a$ (with $a_{\0} = 1$), for the components of a unified background fluid (effective GCG): mass varying dark matter (DM) $\rho_{\sss}$ and cosmon-{like} dark energy (DE).
Three cases are considered for the dynamical neutrino masses: $\mu\bb{\phi} = \phi \exp{\left[-3\bb{\alpha + 1}\phi/2\right]}$ (Case 01),  $\mu\bb{\phi} = \exp{\left[-3\bb{\alpha + 1}\phi/2\right]}$ (Case 02) and $\mu\bb{\phi} = \tanh{\left[3\bb{\alpha + 1}\phi/2\right]}^{\frac{\2\alpha}{\alpha + 1}}\exp{\left[-3\bb{\alpha + 1}\phi\right]}$ (Case 03).
The dark matter densities are computed for $\mathcal{M}\bb{\phi}$ in correspondence to the GCG scenario through Eq.~(\ref{aprl10}), with $A_{\s}= 4/5$ and $\alpha = 1$ (solid line), $1/2$ (dot line), $1/4$ (dash dot line), and $1/8$ (dash line).}
\label{AAA}
\end{figure}

\begin{figure}
\vspace{-3 cm}
\centerline{\psfig{file= 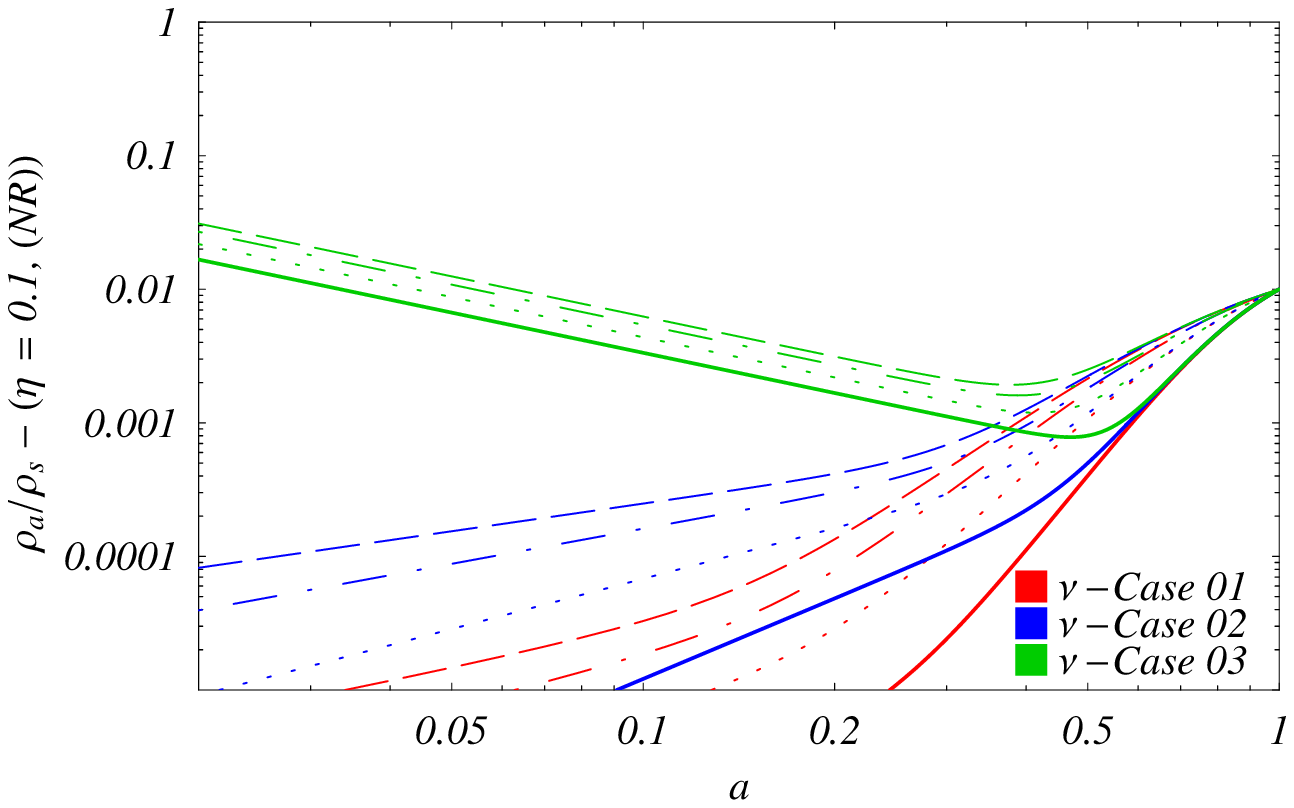,width=9.3 cm}}
\vspace{-3 cm}
\centerline{\psfig{file= 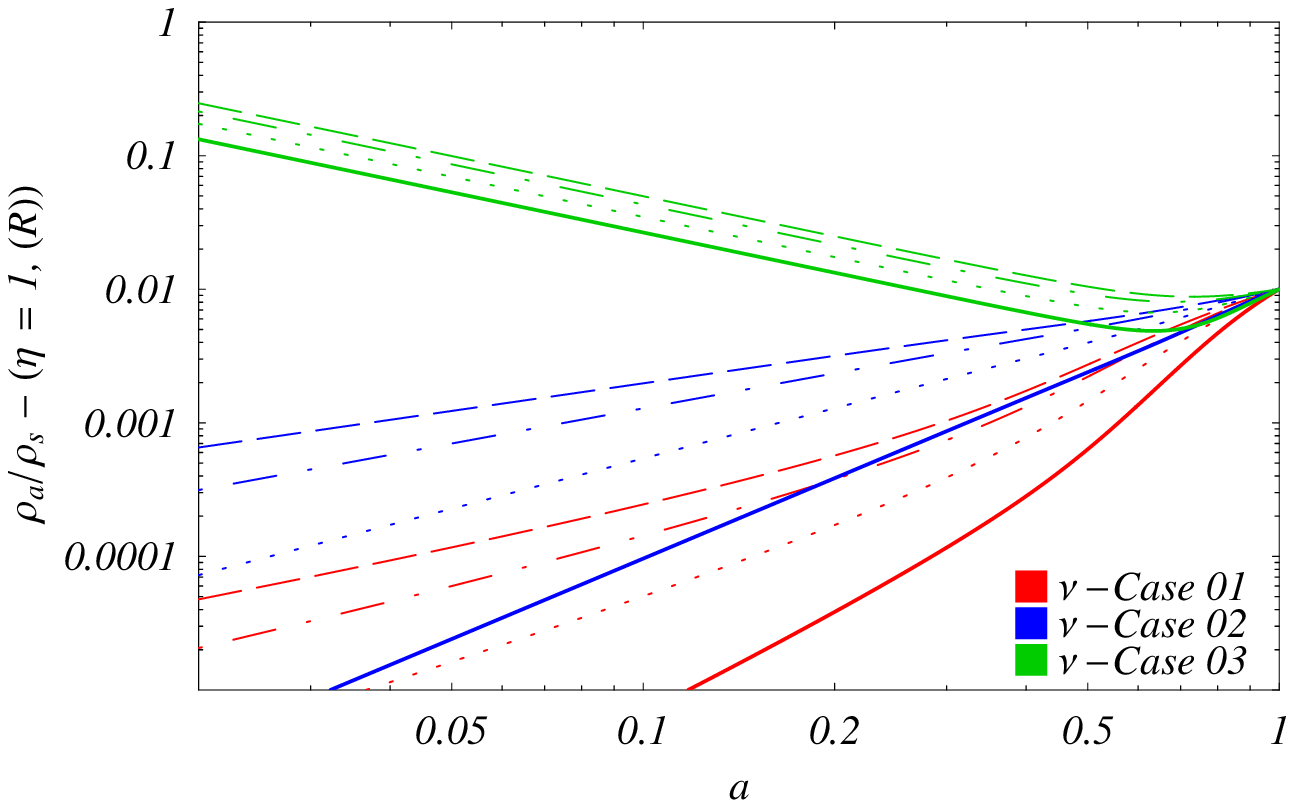,width=9.3 cm}}
\vspace{-3 cm}
\centerline{\psfig{file= 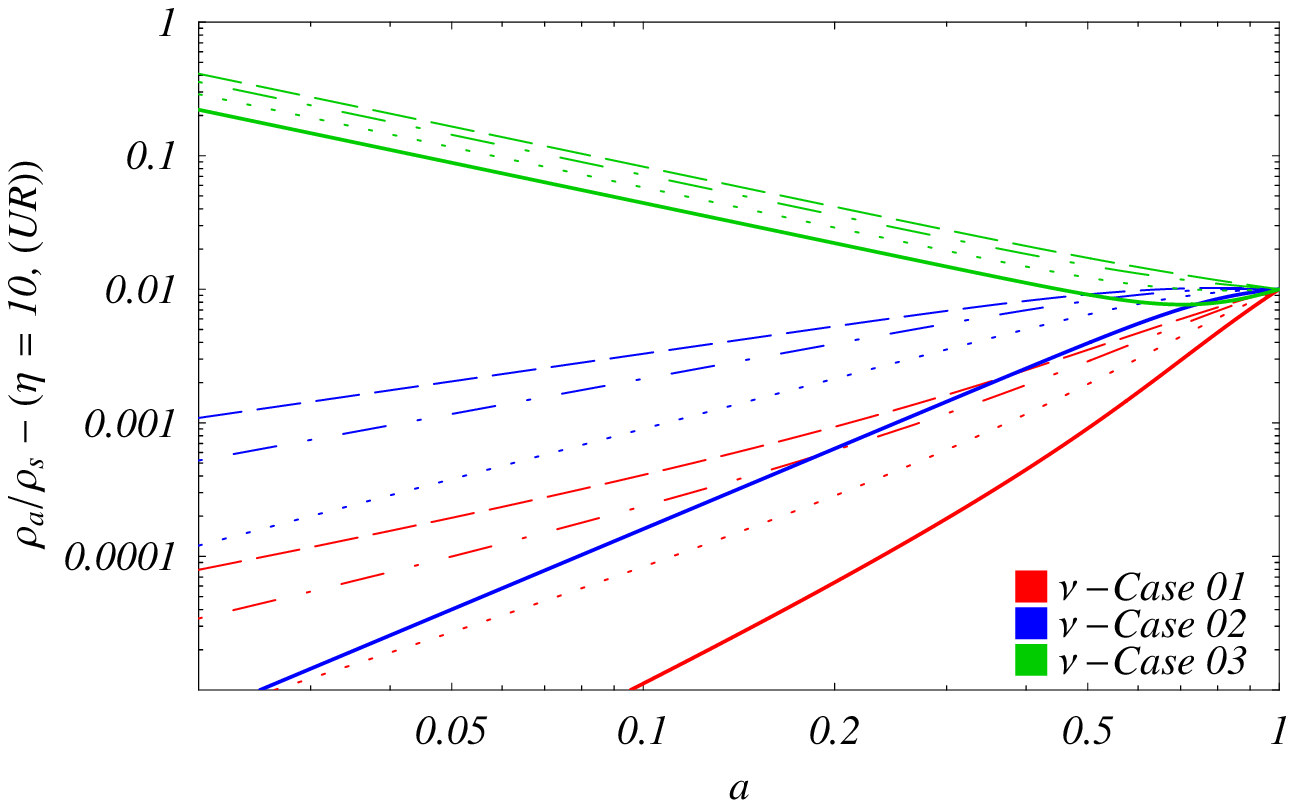,width=9.3 cm}}
\vspace{-2 cm}
\caption{\small Active and sterile neutrino energy density rate $\rho_{\A}/\rho{\sss}$ as function of the scale factor, $a$ (with $a_{\0} = 1$).
Both sterile and active components are assumed to behave like a DFG.
The results are obtained for different GCG models with $A_{\s}= 4/5$ and parameters $\alpha = 1$ (solid line), $1/2$ (dot line), $1/4$ (dash dot line), and $1/8$ (dash line).
Three cases are considered for the neutrino masses (Case 01: red lines, Case 02: blue lines, Case 03: green lines) in agreement with Fig.~\ref{AAA} for nonrelativistic (NR), relativistic and ultrarelativistic (UR) neutrinos at present.}
\label{BBB}
\end{figure}

Considering that the cosmological background evolves according to the GCG model, Fig.~\ref{AAA} exhibits the behavior of the dark sector for the three considered cases.
The evolving neutrino masses lead to evolution of the neutrino energy densities which are shown in Fig~\ref{BBB}, where active and sterile neutrino energy density ratios, $\rho_{\A}/\rho{\sss}$, are plotted as function of the scale factor $a$.
The sterile neutrino is assumed to behave like dark matter.
Notice that different thermodynamic regimes are considered at present: NR, relativistic and UR.
As expected, once the active neutrino reaches the NR regime, the masses and densities are not independent and follow the exponential behavior given by $\rho_{\A}/\rho_{\sss} \propto s^{\2} = \exp{\left[3\bb{\alpha + 1}\phi\right]}$.

\begin{figure}
\centerline{\psfig{file= 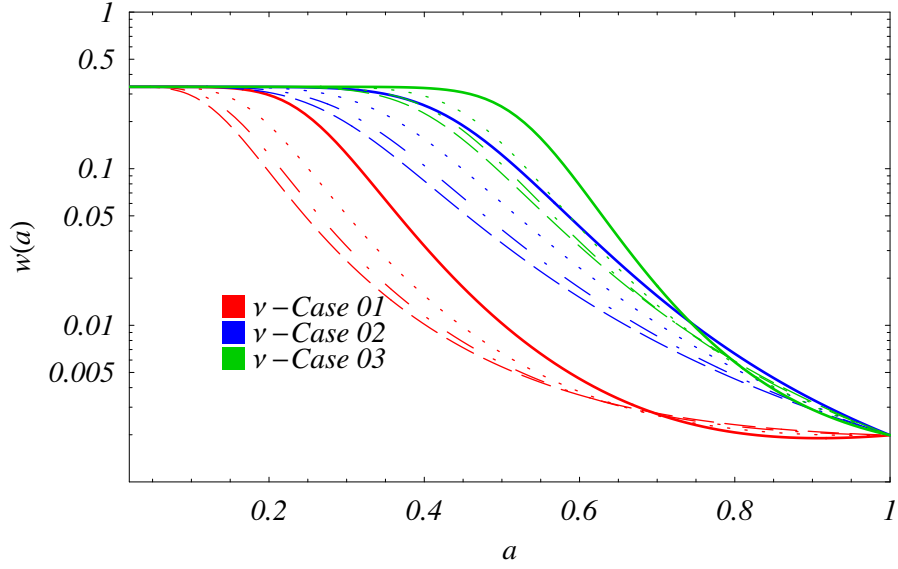,width=13 cm}}
\caption{\small Neutrino equation of state $w = p_{\A}/\rho_{\A}$ for the GCG cosmological model with neutrinos behaving like a DFG in the NR regime ($\eta = 0.1$ at present).
The GCG parameters and the neutrino mass dependence on $\phi$ are in correspondence with the previous figures.}
\label{CCC}
\end{figure}

\begin{figure}
\centerline{\psfig{file= 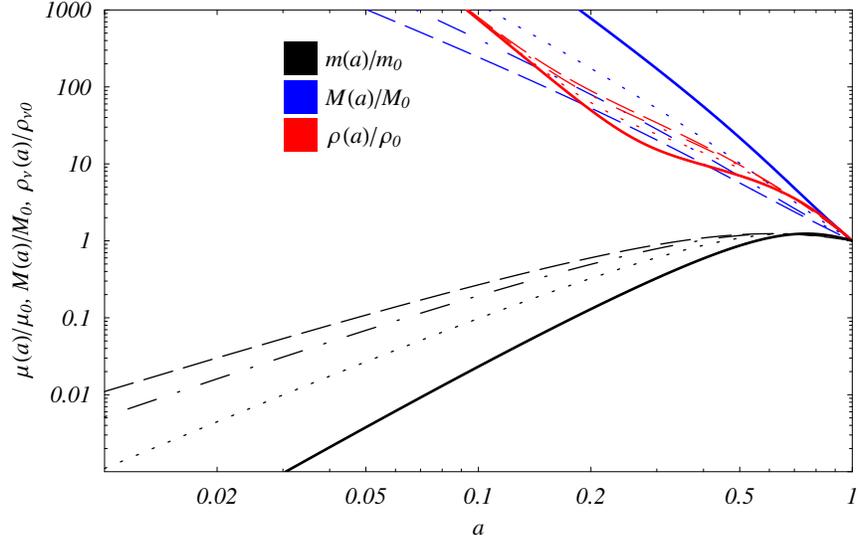,width=13 cm}}
\caption{\small Growing neutrino mass ($\mu\bb{a}/\mu_{\0}$) in dependence on the scale factor, $a$.
Masses are normalized in terms of $\mu_{\0}$ at present ($a = a_{\0} = 1$).
The GCG parameters and the neutrino mass dependence on $\phi$ are in correspondence with the ones in the previous figures.
The GCG scenarios where $A_{\s}= 4/5$ and $\alpha = 1,\,1/2,\,1/4,\,1/8$ are considered.}
\label{DDD}
\end{figure}

The equation of state for active neutrinos in NR regime at present is shown in Fig.~\ref{CCC}.
Despite the explicit model dependence of the masses on the scalar field, Figs.~\ref{BBB} and \ref{CCC} show similar global features that are typical of the proposed framework.
The resulting mass dependence on the scale factor for the three considered cases is depicted in Fig.~\ref{DDD}.
In the three cases one sees a sharp increase of the neutrino masses at a recent past ($a \gtrsim 0.2$).

Our analysis shows that the active neutrino mass component grows when coupled to dark matter through the {\em seesaw cosmon} field within the GCG model.
This is consistent with the treatment of neutrinos as a test fluid, in the sense that its cosmological equation of motion decouples from the dark sector which governs the cosmological evolution.
In addition, it provides the conditions to analyze instabilities.
The possibility of adiabatic instabilities in cosmological scenarios was previously pointed out in the context of a mass varying neutrino model of dark energy whereas the dynamical dark energy model is obtained through the coupling of a light scalar field to neutrinos, but not to dark matter \cite{Afs05}.
The ensued effects have been extensively discussed in the context of mass varying neutrinos, in which the growth of the neutrino's mass and the recent accelerated expansion are linked through the scalar field coupling.
In the adiabatic regime, these models face catastrophic instabilities on small scales, since the sound speed squared of the coupled fluid is negative.
Starting with a uniform fluid, such instabilities would give rise to exponential growth of small perturbations.
The natural interpretation of this is that the universe becomes inhomogeneous with neutrino overdensities subject to nonlinear fluctuations which eventually collapse into compact localized regions \cite{Mot08}.

Our approach allows one to perform a stability analysis for NR active neutrinos independently of the dark sector evolution.
The first step is obtaining the squared speed of sound, $c^{\2}_{s}$, for neutrinos.

\begin{figure}
\vspace{-3 cm}
\centerline{\psfig{file= 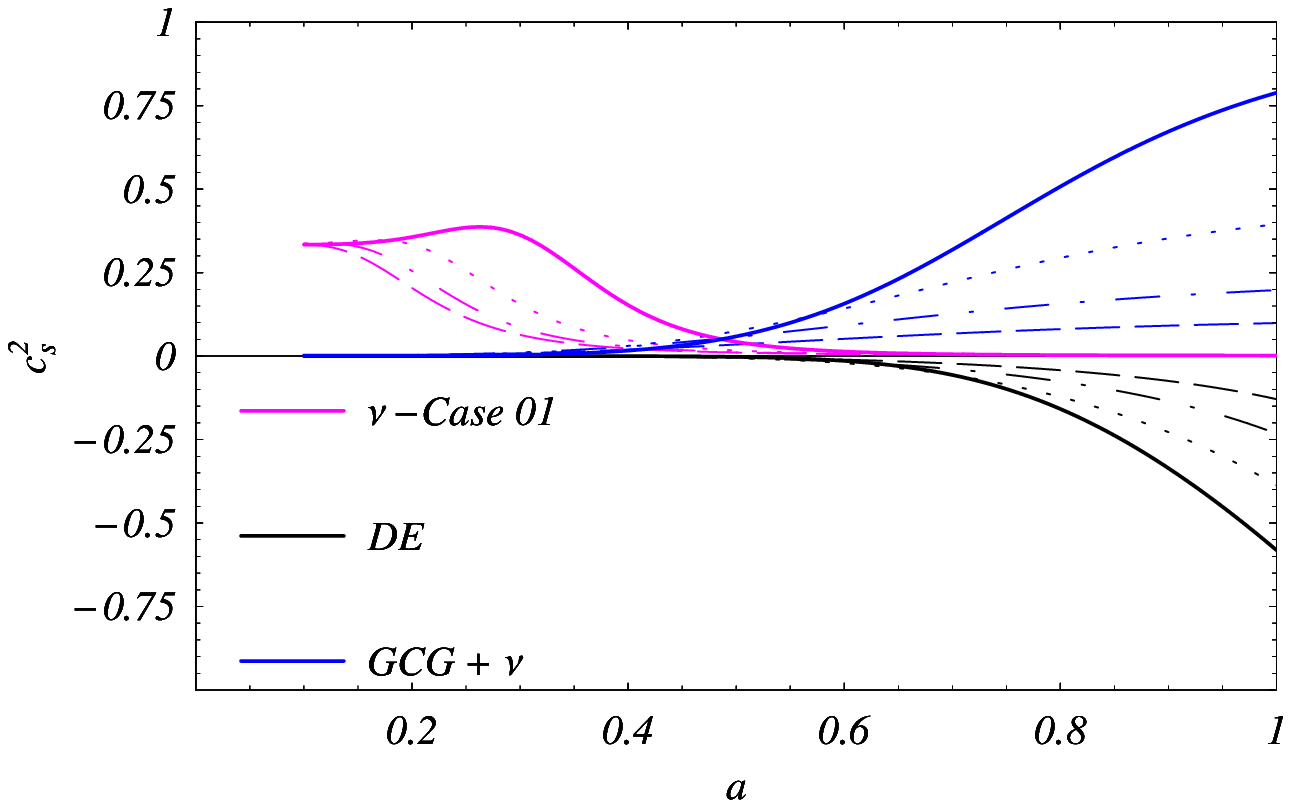,width=10 cm}}
\vspace{-3 cm}
\centerline{\psfig{file= 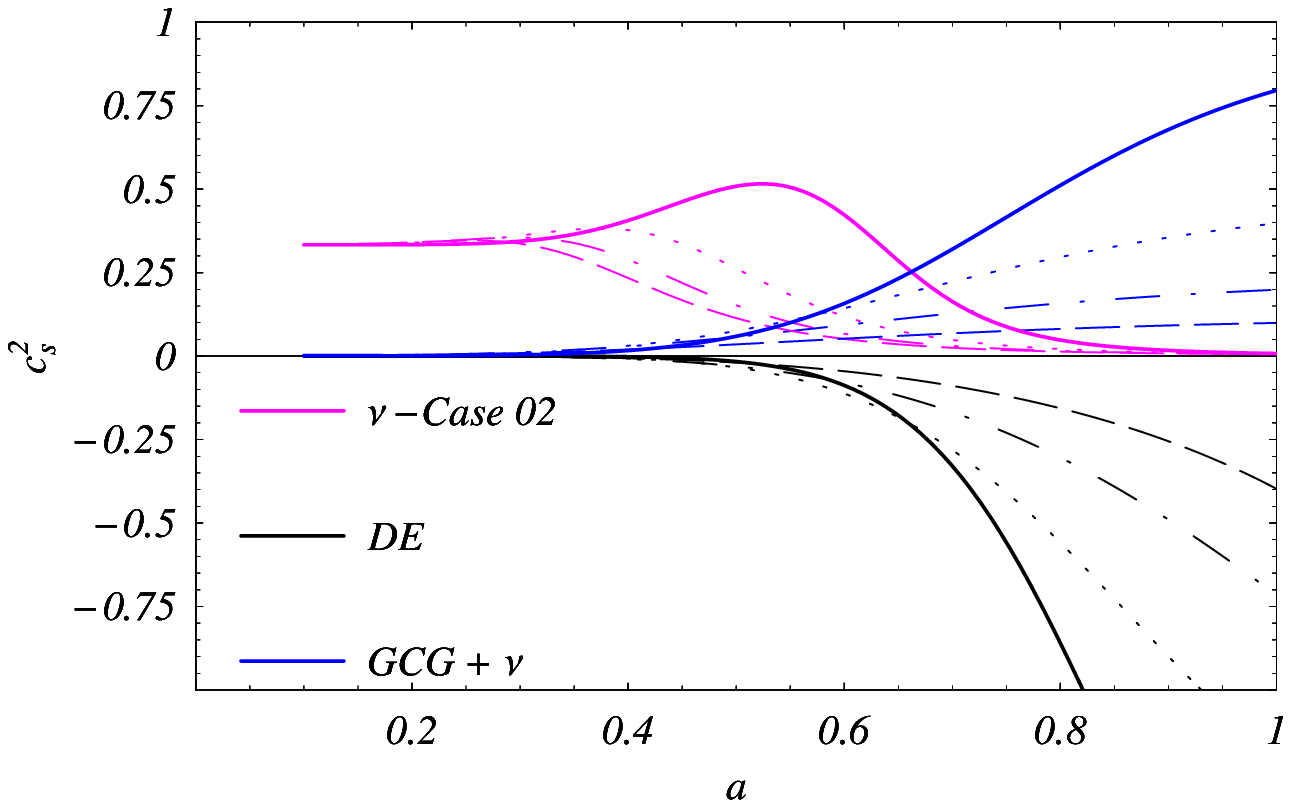,width=10 cm}}
\vspace{-3 cm}
\centerline{\psfig{file= 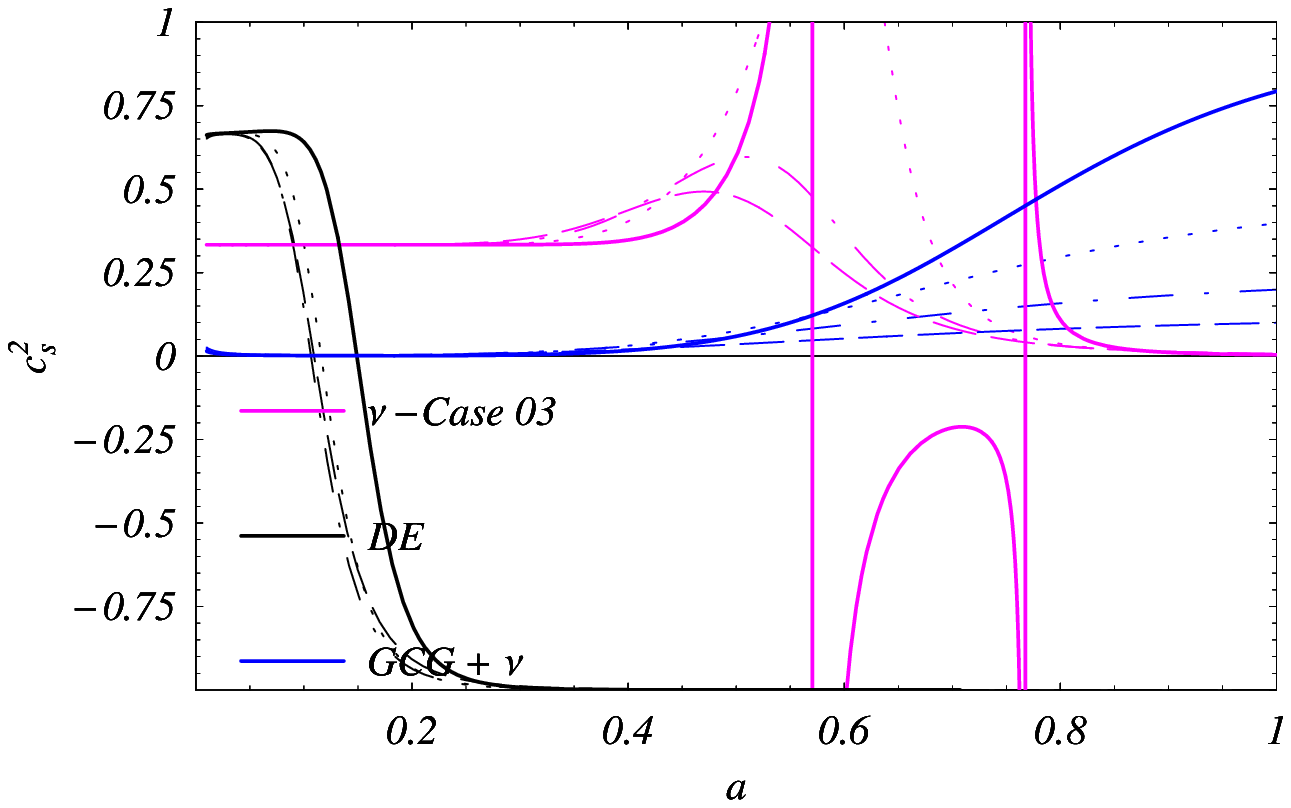,width=10 cm}}
\vspace{-2 cm}
\caption{\small Squared speed of sound $c^{\2}_{s} = \mbox{d}p/\mbox{d}\rho$ in dependence on the scale factor, $a$ (with $a_{\0} = 1$), for neutrinos (magenta lines), for the cosmon-{\em like} dark energy component - DE (black lines) and for the corresponding GCG scenario (blue lines).
The GCG parameters and the neutrino mass dependence on $\phi$ are in correspondence with the previous figures.
Neutrinos are assumed to behave like a DFG in the NR regime ($\eta = 0.1$ at present).}
\label{EEE}
\end{figure}

Fig.~\ref{EEE} illustrates the results for $c^{\2}_{s}$ for DFG of neutrinos ($\nu$), the {\em seesaw cosmon} dark energy and the GCG plus neutrinos.
The results are compared with those of a GCG scenario with $A_{\s}= 4/5$ for $\alpha = 1$ (solid line), $1/2$ (dot line), $1/4$ (dash dot line), and $1/8$ (dash line).
The role of neutrinos on the positiveness of $c^{\2}_{s}$ is measured in terms of those for the GCG scenario.
One sees that the influence of neutrinos on the positiveness of $c^{\2}_{s}$ for the GCG is not relevant.
Actually for the first two cases, NR neutrinos lead to vanishing perturbations on the background at present, so that the test fluid approach is quite accurate.

The results obtained from the analysis of the third case show that for the models where the stationary condition (cf. Eq.~(\ref{pap09})) implies that, $ p_{\phi} = - \rho_{\phi}$, one obtains $c_{s}^{\2} = -1$ from the very start.
The effective GCG plus neutrinos is free from this inconsistency, i. e. the coupling of the dark energy component with mass varying dark matter allows for removing such inconsistency as $c_{s}^{\2} \simeq d p_{\phi}/ d\rho_{\phi} > 0$.

\section{Conclusions}

In this work it is shown that GCG contains some key ingredients in the description of the universe dynamics and that it allows for a suitable background for the mass varying and the {\em seesaw} mechanisms.
Furthermore it has been pointed out that the mass varying behavior of the dark matter component can be matched with the GCG model.

The main features of the proposed model can be summarized as follows:
i) The cosmological evolution of neutrino energy densities is driven by coupled {\em cosmon}-type field equations where active and sterile neutrino states are connected through the {\em seesaw} mechanism for mass generation.
ii) Dark matter is, most often, not considered in the MaVaN models.
The treatment of dark energy and dark matter in the GCG unified scheme naturally offers this possibility.
Identifying sterile neutrinos as dark matter coupled with dark energy provides the conditions to implement such unified picture in the MaVaN formulation.
Moreover, the constraints imposed by the {\em seesaw} mechanism in order to establish a unique analytical connection to the GCG in terms of a real scalar field were found.
iii) The dynamics of the coupled fluid composed by neutrinos, dark matter and dark energy is driven by one single degree of freedom, the scalar field, $\phi\bb{a}$.
Since the GCG allows for an explicit dependence on the scale factor, $a$, $s$ and $\xi$ dependence on the scalar field do imply a dependence on the universe's evolution.
Due to the connection between the GCG and the {\em seesaw} masses, the proposed approach actually yields an effective model for MaVaN's coupling to the dark sector.
Of course, one can interpret the mediation of the scalar field as a dependence on the dimensionless scales $a/a_{\0}$, $s$, and $\xi/M$.

The described mechanism leads to a fast increase of the neutrino masses, which results in a model dependent vanishing speed of sound at present.
The dynamical mass behavior, due to the evolution of the {\em seesaw cosmon} field, coincides in a subtle way with the GCG dynamics.
However, this scenario is not unique.
Nevertheless, without any additional assumptions, it provides an attractive solution for the coincidence problem through the confluence of three independent frameworks: the {\em cosmon}-like dynamics, the {\em seesaw} mechanism for mass generation and the GCG model.
For active neutrinos, an increase of the mass $\mu$ by a factor $10^{\6}$, corresponds approximately to a decrease of the sterile neutrino mass $\mathcal{M}$ from the Planck scale to $10^{\1\3}\, GeV$.
Thus, the GCG model modulates the increase of the neutrino mass, which acts as a cosmological clock to the present scenario of cosmological accelerated expansion.

Unfortunately, in general, one cannot provide a sharp criterion for the potential and for the mass varying dependence on the scalar field to discriminate between the GCG scenario and a unified fluid via a {\em seesaw cosmon} field that mimics the GCG.
Furthermore, it has been shown for specific potentials that, in what concerns stability and cosmic acceleration, many results found in the literature can be recovered.
Actually, new scenarios featuring other mass dependencies on the {\em seesaw cosmon} field can also be considered.

To conclude, the proposed unified scheme, despite similarities with some quintessence models, is an ambitious and encompassing scheme where mass varying particles can be related with the stability and the cosmic acceleration of the universe.

\begin{acknowledgments}
A. E. B. would like to thank for the financial support from the Brazilian Agencies FAPESP (grant 08/50671-0) and CNPq (grant 300627/2007-6).
\end{acknowledgments}

\section*{Appendix: The mass varying mechanism for a DFG}

Aiming to understand how the mass varying behavior takes place, it is assumed that for a particle statistical distribution $f\bb{q}$, where $q \equiv \frac{|\mbox{\boldmath$p$}|}{T_{\0}}$, $T_{\0}$ being the background temperature at present, in a flat FRW cosmological scenario.
For a generic dynamical mass defined by $\mbox{\textsf{M}}\bb{\phi}$, the corresponding particle density, energy density and pressure are expressed as
\begin{eqnarray}
n\bb{a} &=&\frac{T^{\3}_{\0}}{\pi^{\2}\,a^{\3}}
\int_{_{0}}^{^{\infty}}{\hspace{-0.3cm}dq\,q^{\2}\ \hspace{-0.1cm}f\bb{q}},\nonumber\\
\rho\bb{a, \phi} &=&\frac{T^{\4}_{\0}}{\pi^{\2}\,a^{\4}}
\int_{_{0}}^{^{\infty}}{\hspace{-0.3cm}dq\,q^{\2}\, \left(q^{\2}+\frac{\mbox{\textsf{M}}^{\2}\bb{\phi}\,a^{\2}}{T^{\2}_{\0}}\right)^{\1/\2}\hspace{-0.1cm}f\bb{q}},\\
p\bb{a, \phi} &=&\frac{T^{\4}_{\0}}{3\pi^{\2}\,a^{\4}}\int_{_{0}}^{^{\infty}}{\hspace{-0.3cm}dq\,q^{\4}\, \left(q^{\2}+\frac{\mbox{\textsf{M}}^{\2}\bb{\phi}\,a^{\2}}{T^{\2}_{\0}}\right)^{\mi\1/\2}\hspace{-0.1cm} f\bb{q}},~~~~ \nonumber
\label{gcg01}
\end{eqnarray}
where one has introduced the sub-index $0$ for denoting present-day values, with $a_{\0} = 1$.

In the limit where $T$ tends to $0$, the Fermi distribution $f\bb{q}$ becomes a step function that yields an elementary integral with the upper limit equal to the Fermi momentum $q_{F}$.
The energy density and pressure of a DFG can be expressed in terms of elementary functions of the scale factor, $\beta \equiv\beta\bb{a} = T_{\0}\,q_{F}/a$ and $\mbox{\textsf{M}} \equiv \mbox{\textsf{M}}\bb{\phi\bb{a}}$,
\begin{eqnarray}
n\bb{a} &=&\frac{1}{3 \pi^{\2}}\beta^{\3},\nonumber\\
\rho\bb{a} &=& \frac{1}{(8 \pi^{\2})}
\left[\beta (2 \beta^{\2} + \mbox{\textsf{M}}^{\2})\sqrt{\beta^{\2} + \mbox{\textsf{M}}^{\2}} -
\mbox{\textsf{M}}^{\4}\,arc\sinh{\left(\beta/\mbox{\textsf{M}}\right)}\right],\\
p\bb{a} &=&\frac{1}{(8 \pi^{\2})}
\left[\beta (\frac{2}{3} \beta^{\2} - \mbox{\textsf{M}}^{\2})\sqrt{\beta^{\2} + \mbox{\textsf{M}}^{\2}} + \mbox{\textsf{M}}^{\4}\,arc\sinh{\left(\beta/\mbox{\textsf{M}}\right)}\right].\nonumber
\label{gcg01B}
\end{eqnarray}
Simple mathematical manipulations allow one to show that
\begin{equation}
n\bb{a} \frac{\partial \rho\bb{a}}{\partial n\bb{a}} = (\rho\bb{a} + p\bb{a}),
\label{gcg02B}
\end{equation}
and
\begin{equation}
\mbox{\textsf{M}}\bb{a} \frac{\partial \rho\bb{a}}{\partial \mbox{\textsf{M}}\bb{a}} = (\rho\bb{a} - 3 p\bb{a}),
\label{gcg02}
\end{equation}
from which one can obtain the energy-momentum conservation equation
\begin{equation}
\dot{\rho} + 3 H (\rho + p) - \dot{\phi}\frac{\mbox{d} \mbox{\textsf{M}}}{\mbox{d} \phi} \frac{\partial \rho}{\partial \mbox{\textsf{M}}} = 0.
\label{gcg03}
\end{equation}
The coupling between relic particles and the scalar field as described by Eq.~(\ref{gcg02}) is relevant only in the NR regime since the strength of the coupling is suppressed by the pressure of the relativistic ($T\bb{a} = T_{\0}/a >> \mbox{\textsf{M}}\bb{\phi\bb{a}}$) particles, so that matter and scalar field fluids tend to decouple and evolve adiabatically.

\pagebreak
\newpage

\end{document}